\documentclass[12pt,preprint]{aastex}
\usepackage{amsmath}
\usepackage[normalem]{ulem} 

\def\HeII{$^3$He\,{\scriptsize II}}

\def\bea{\begin{eqnarray}}
\def\eea{\end{eqnarray}}
\def\be{\begin{equation}}
\def\ee{\end{equation}}

\newcommand{\showlabel}[1]{} 
\newcommand{\internalcom}[1]{}

\received{}
\accepted{}
\journalid{}{}
\articleid{}{}
\paperid{}
\cpright{AAS}{2011}
\ccc{}

\shorttitle{Hyperfine-changing transition in one-electron ions}
\shortauthors{Bartschat and Sadeghpour}

\begin{document}

\title{Hyperfine-changing transitions in $^3$He\,{\normalsize II}\ and other one-electron ions by electron scattering}

\author{Klaus Bartschat\altaffilmark{1}}
\affil{Department of Physics and Astronomy, Drake University, Des Moines, Iowa, 50311, USA}
\email{klaus.bartschat@drake.edu}

\and

\author{H. R. Sadeghpour}
\affil{ITAMP, Harvard-Smithsonian Center for Astrophysics, Cambridge, MA 02138, USA}
\email{hrs@cfa.harvard.edu}

\altaffiltext{1}{ITAMP, Harvard-Smithsonian Center for Astrophysics, Cambridge, MA 02138, USA}


\begin{abstract} 
We consider the spin-exchange (SE) cross section in electron scattering from $^3$He\,{\scriptsize II}, which drives the
hyperfine-changing  \hbox{3.46~cm} (8.665~GHz) line transition.
Both the analytical quantum defect method --- 
applicable at very low energies --- and accurate \hbox{R-matrix} techniques for electron-He$^+$ 
scattering are employed to obtain SE cross sections. 
The quantum defect theory is also applied to electron collisions with other one-electron ions in order
to demonstrate the utility of the method and derive scaling relations. 
At very low energies, the hyperfine-changing cross sections due to e$-$He$^+$ scattering are much 
larger in magnitude than for electron collisions with neutral hydrogen, hinting at large rate constants 
for equilibration. Specifically, we obtain rate coefficients of 
$K(10\,{\rm K}) = 1.10 \times 10^{-6}\,\rm cm^3/s$ 
and $K(100\,{\rm K}) = 3.49\times 10^{-7}\,\rm cm^3/s$.
\end{abstract}


\section{Introduction}
The signature of matter formation, the epoch of recombination, is recorded on the 
relic radiation, the cosmic microwave background (CMB). This imprint occurred 
at the surface of last scattering, corresponding to a redshift of $z\approx 1100$. 
Stars ionized matter much later, not until $6 < z< 30$ \citep{barkana2001}.  
Observations of Lyman-$\alpha$ (Ly-$\alpha$) absorption at $z\approx 6$ in 
quasars have pinned the end of the reionization epoch \citep{fan2002}. 
This epoch, called the cosmic reionization, is frontier astronomy, because 
there are few observational constraints on the nature 
of processes during this era. 
A possible relic radiation from the reionization 
era could come from the hydrogen hyperfine \hbox{21-cm} (1.420 GHz) tomography \citep{loeb2004}, 
when the spin temperature of the \hbox{21-cm} line ($T_s$) \citep{purcell1956} falls below that of the 
CMB radiation ($T_{\rm cmb}$), rendering the high redshift \hbox{21-cm} signal visible in the CMB. 

The hyperfine transition in atomic hydrogen, and also in singly-ionized He (\HeII), occurs through 
the $F=1 \to 0$ transition. This spin-changing transition can occur either via Raman 
scattering of red-shifted stellar ultra\-violet light (into Ly-$\alpha$ resonance, the 
so-called Wouthuysen-Field effect \citep{barkana2005}), by collisions due to {\bf spin dipole-dipole interaction}, or by electron scattering. 
The process of radiation scattering brings the photons into Boltzmann equilibrium 
with the surrounding gas, and hence the spin temperature into equilibrium with the 
gas kinetic temperature, $T_k$. The equilibration defines a unique spin temperature. 
It has been shown that collisional dipolar {\bf processes \citep{zygelman2005} are}
ineffective in the redistribution of hyperfine level populations into statistical 
equilibrium. On the other hand, {\bf electron-induced SE cross 
sections can be orders of magnitude larger than those resulting from dipolar collisions}
and can readily lead to equilibration.

The hyperfine-changing transition in $^3$He$^+$ at 3.46~cm (8.665~GHz) 
is used for precise determination of the $^3$He/H abundance in the 
interstellar medium \citep{bania1997}. The source of ionization includes 
\hbox{H\,{\scriptsize II}} regions and planetary nebulae. $^3$He can be a proxy for probe of 
cosmology as well as stellar and galactic evolution. In models to obtain the $^3$He/H ratio, 
care must be taken to include non-LTE (\hbox{local} thermal equilibrium) effects and 
line broadening due to electron collisions \citep{balser1995}. As will be demonstrated below, 
low-energy SE collision 
of e$^-$-\HeII\ has a much larger cross section 
than SE collision of e$^-$-hydrogen or neutral-neutral hydrogen. 
This is due to the presence of the long-range Coulomb interaction, which changes
the energy dependence of the cross section substantially.

Spin-exchange effects in highly charged ions are also of interest for
many storage ring experiments.  Calculations using the Coulomb-Born and
static exchange approximation, both in non\-relativistic and relativistic
form, were carried out by \citet{AMG1989}. 
They focused on the angle-integrated spin-exchange 
cross section \citep{DR1965}
\begin{equation}
\sigma_{\rm SE} = \frac{\pi}{k^2} \sum_{\ell=0}^\infty 
    (2\ell +1) \sin^2(\delta_\ell^{t} - \delta_\ell^{s})
    \label{eq:sigmaex}
\end{equation}
for elastic collisions, where
$k$ and $\ell$ denote the electron's linear and angular momenta while the superscripts
on the short-range potential phase shifts, $\delta_\ell^{t}$ and $\delta_\ell^{s}$, 
refer to triplet ($t$) and singlet~($s$) 
scattering with total electron spin $S=1$ or $S=0$, respectively.
For the above definition to make physical sense, the HFS splitting 
(about $3.5 \times 10^{-5}$~eV, corresponding to a temperature of about 0.4~K) 
should be small compared to the collision energy. {\bf For the purposes intended here, this condition is well fulfilled.}

In this work, we calculate the SE cross sections for scattering of 
electrons from He$^+$ and other one-electron ions with low to intermediate nuclear
charges of $2 \le Z \le 26$.  We treat the problem in a non\-relativistic 
framework, but for this formulation we employ much more sophisticated 
methods of treating the collision process, including a convergent 
\hbox{R-matrix} (close-coupling) with pseudo\-states (RMPS) model that is expected
to provide cross sections with an uncertainty of a few percent (in the worst 
scenario) for the energy range considered here. 
Unless specified otherwise, atomic units are used throughout this paper.
 
\section{Numerical Methods}
The collision calculations for electron scattering from He$^+$ were performed
using the \hbox{R-matrix} approach \citep{Burke2011} as a means to solve the resulting
close-coupling equations.  We performed one set of calculations in a simple three-state model
(to be referred to as \hbox{RM-3} below), 
only coupling the $\rm (1s)^2S$, $\rm(2s)^2S$, and $\rm(2s)^2S$ states of He$^+$.
This model was then extended to a 23-state RMPS approach (\hbox{RM-23}) 
based on the general method outlined in \citet{BHSBB1996}. Here we added
the physical $n=3$ states plus a number of pseudo\-states to approximate the
effect of the higher-lying Rydberg states as well as the ionization continuum on
the numerical predictions.  This model has been tested in great detail on several 
occasions, including the description of the initial bound state and the 
ejected-electron$-$residual-ion interaction in calculations of electron-impact 
ionization \citep{PRL2005,PRL2006,PRA2014}.  

It is also possible to study SE in electron$-$ion collisions by invoking 
Quantum Defect Theory (QDT). Specifically, the
quantum defect, i.e., the difference~$\mu_\ell \equiv n - n_\ell^*$ between 
the effective quantum number~$n_\ell^*$ and the nominal quantum number~$n$ 
in the Rydberg formula for the binding energies,
\begin{equation}
    \label{eq:energy}
   E_{nl} = -\frac{1}{2} \frac{Z^2}{n_\ell^*},
\end{equation}
is a measurement for the deviation from the pure Coulomb potential of a nuclear
charge $Z-1$ due to the imperfect screening by the inner $1s$ electron.  
For $n \to \infty$, the quantum defect will converge to $\bar{\mu}_\ell$, which is
related to the potential scattering phase shift $\delta_\ell$ at the elastic
threshold of zero energy via~\citep{QDT}
\begin{equation}
   \label{eq:phase}
   \lim_{E \to 0} \delta_\ell(E) = \pi \bar{\mu}_\ell.
\end{equation}
For the practical purposes of interest here, the quantum defect is sufficiently 
converged by $n \approx 6$, and we will see
that the energy dependence of the phase shift is smooth over a relatively large 
energy range -- except, of course, when resonances will come into play.   Consequently,
if accurate energy levels are available (preferably taken from experiment, but a
reasonably sophisticated theory would also be appropriate), one can
estimate the spin-exchange cross section to a high degree of accuracy without actually
performing an extensive numerical calculation.  In fact, as will be illustrated 
in the next section, scaling relations and the fast convergence of the
results can be explained in a straightforward way using QDT arguments.

\section{Results}

\begin{figure}[t]
\centerline{\includegraphics[width=0.80\textwidth]{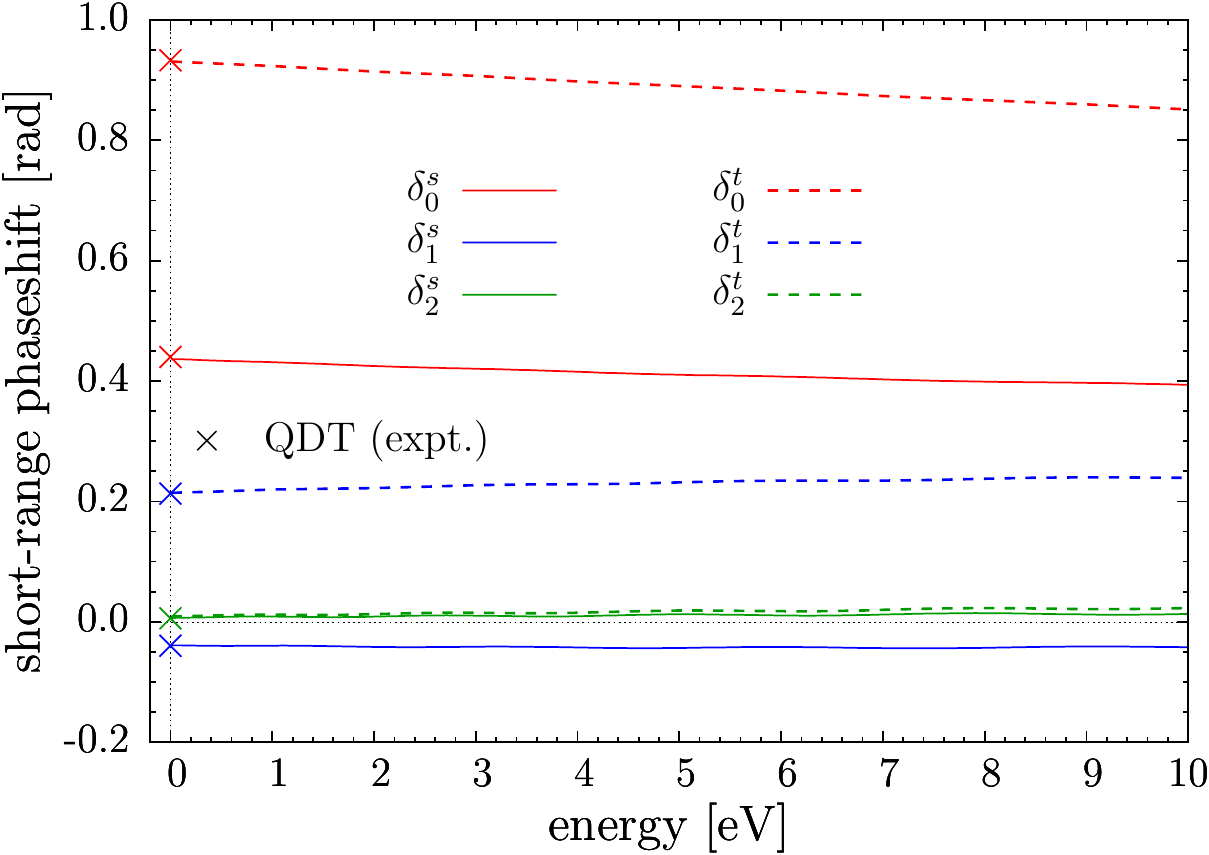}}
\caption{\label{fig:fig1}  Potential phase shifts for elastic 
                           electron scattering from He$^+$ from a 23-state RMPS calculation.
                           There is virtually perfect agreement with what one would expect from
                           quantum defect theory (indicated by the crosses at $E = 0$).
                           The quantum defects were estimated from the energy levels
                           of the NIST database \citep{NIST}.
                           See the electronic edition of the Journal for a color version of this figure.}
\end{figure}

Figure~\ref{fig:fig1} shows the short-range potential phase shifts from a 23-state RMPS 
calculation. Note the virtually perfect agreement with what one would expect from
QDT at zero energy; there is only a very weak dependence on the electron energy.  Also, 
due to the centri\-fugal barrier, only partial waves with small angular momentum
(in practice $\ell = 0$ and $\ell = 1$) will ``see'' a significant deviation of the
actual potential from the ideal $Z-1$ Coulomb potential, where one of the
nuclear charges has been fully screened by the inner $1s$ electron. 

\begin{figure}[t]
\centerline{\includegraphics[width=0.80\textwidth]{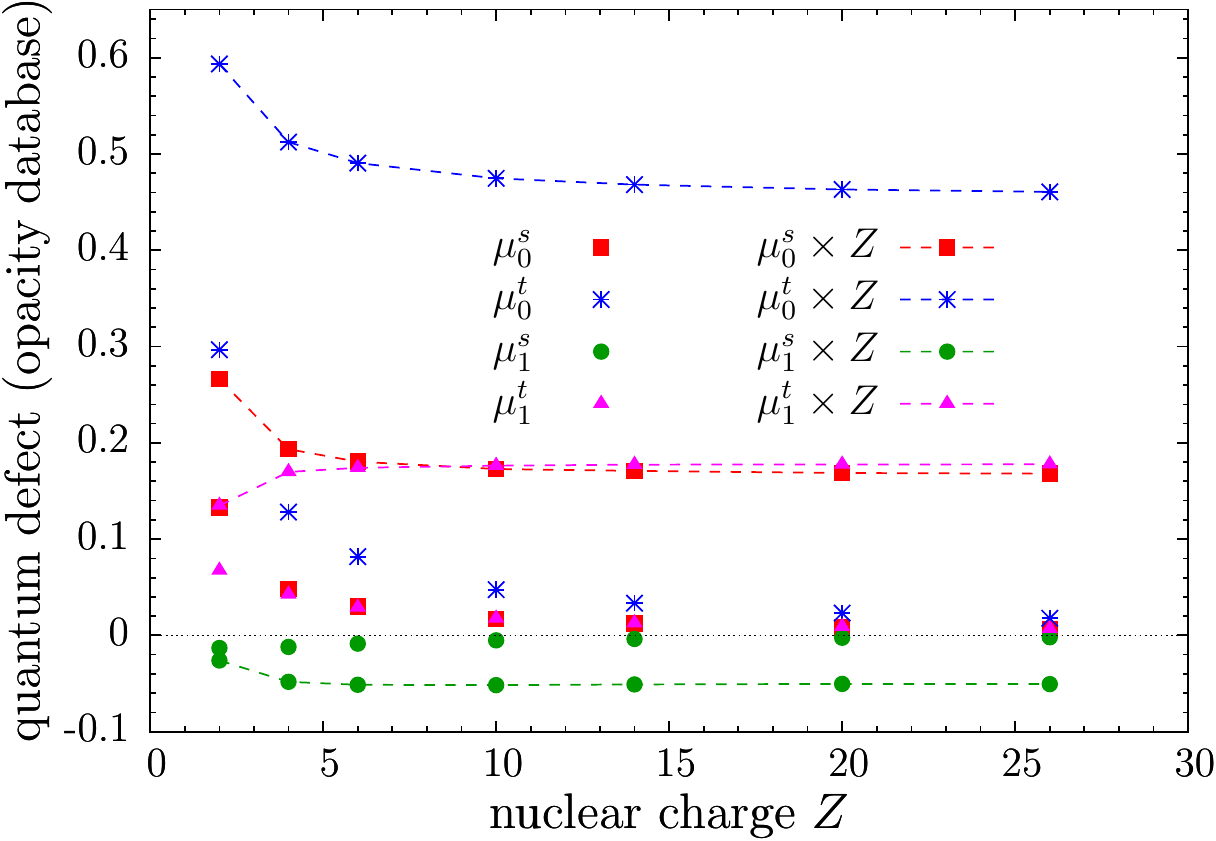}}
\caption{\label{fig:fig2} Quantum defects for two-electron systems as a function
of the nuclear charge~$Z$, based on the database of the Opacity Project \citep{OP}.  
The quantum defects scale as $1/Z$, as one would expect, with a slight deviation for small~$Z$.
See the electronic edition of the Journal for a color version of this figure.}
\end{figure}

Figure~\ref{fig:fig2} shows the quantum defects for two-electron systems as a function
of the nuclear charge~$Z$, based on the database of the Opacity Project \citep{OP}.  
As mentioned above, the quantum defect is a measure of imperfect screening due to the
inner $1s$ electron.  Consequently, one would expect this effect to be inversely proportional
to the nuclear~$Z$, except perhaps for relatively small~$Z$.  As seen from the figure, this
expectation is very well fulfilled.  It then follows from Eq.~(\ref{eq:phase}) that the
phase shifts should also fall off like $1/Z$ with increasing~$Z$, and hence one
would expect the SE cross section to scale like~$1/Z^2$.  This scaling relation was 
derived by \citet{AMG1989} based on the properties of their
specific numerical model, but we note here that this is a general feature grounded in the under\-lying physics.

\begin{figure}[t]
\centerline{\includegraphics[width=0.80\textwidth]{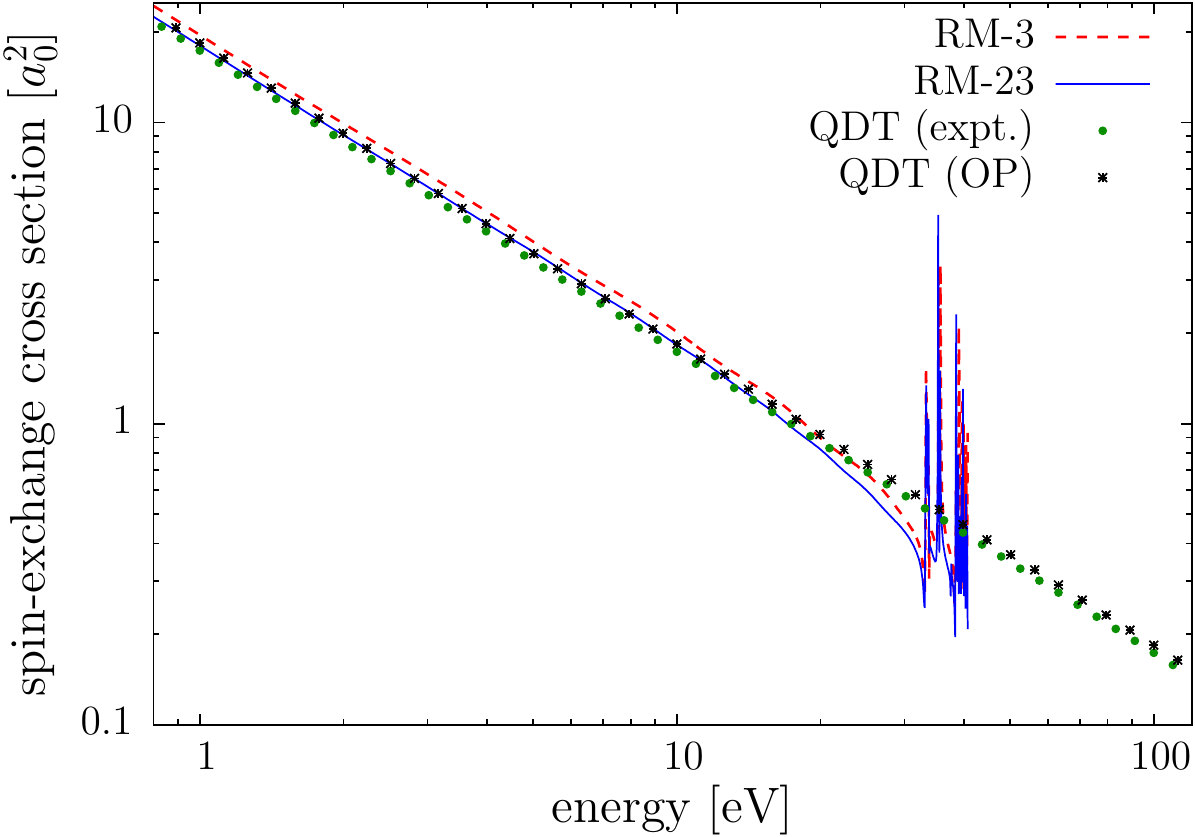}}
\caption{\label{fig:fig3} Spin-exchange cross section for e$-$He$^+$ scattering
as a function of the electron energy.  Numerical results from a 23-state RMPS model are compared
with predictions from a much simpler 3-state approach, as well as QDT results obtained with
quantum defects from experiment \citep{NIST} or the opacity database \citep{OP}. The inverse linear dependence with energy of the SE cross sections, as predicted from the QDT analysis is confirmed in the numerical calculations. 
See the electronic edition of the Journal for a color version of this figure.}

\end{figure}

Figure~\ref{fig:fig3} shows the spin-exchange cross section for e-He$^+$ scattering
as a function of the electron energy.  Not surprisingly, the
numerical results from the 23-state RMPS model are in excellent agreement with
the QDT predictions, even though the latter are, in principle, only valid at zero energy.
They can, of course, not explain the strong energy dependence due to the series of
Rydberg resonances starting at energies around 20~eV.
The calculations were performed for angular momenta up to $\ell = 4$, but they are converged -- to the
thickness of the line -- by only accounting for $\ell = 0$ and $\ell =1$. As already mentioned
above, this fast convergence of the partial-wave expansion
is a straight\-forward consequence of the centri\-fugal barrier, which leads to 
small quantum defects and, therefore, small short-range potential scattering phases.  
The long-range Coulomb phase,
responsible for an infinite total cross section for elastic electron$-$ion scattering, is
spin independent, and hence it cancels out when the difference in the spin phase shifts is taken. 

\begin{figure}[t]
\centerline{\includegraphics[width=0.80\textwidth]{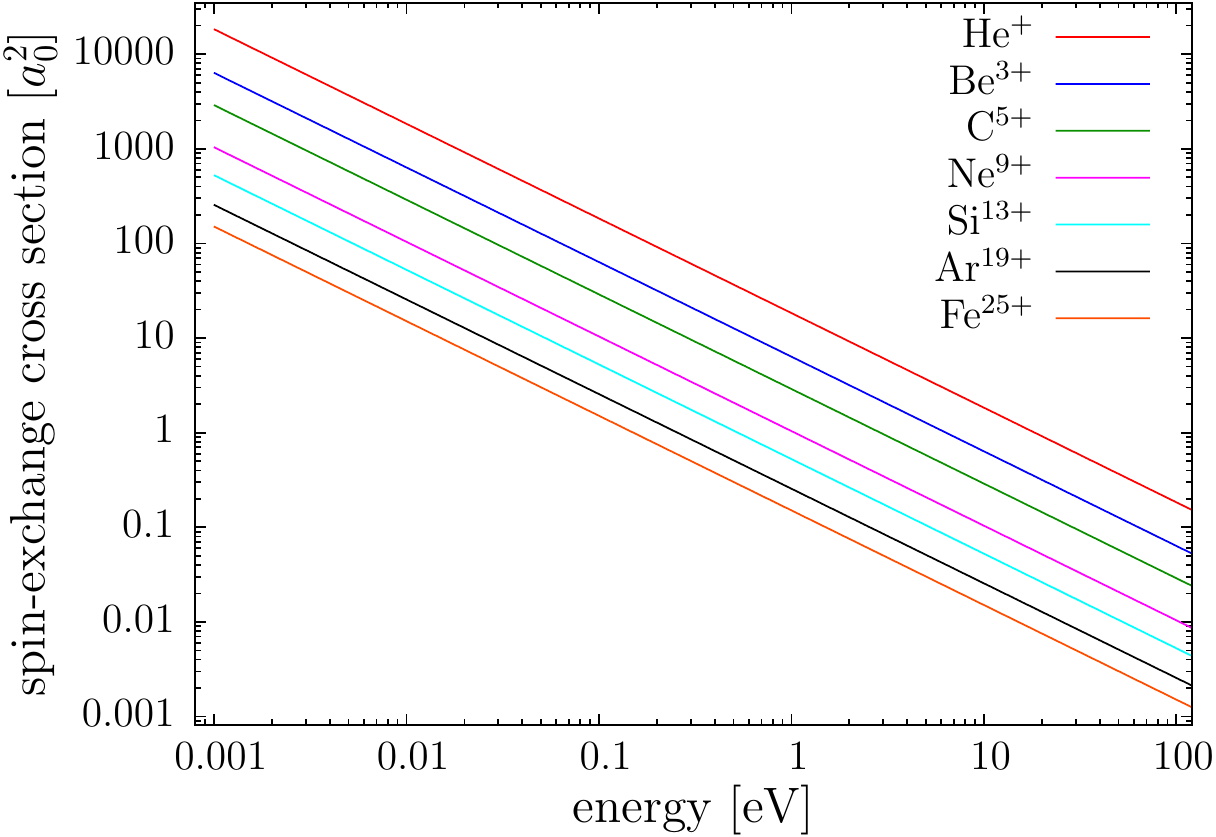}}
\caption{\label{fig:fig4} Spin-exchange cross sections predicted from
QDT results obtained with quantum defects from the Opacity Project database \citep{OP}.
See the electronic edition of the Journal for a color version of this figure.}
\end{figure}

Once again considering practical applications, we notice that the  results from the 
three-state model already appear to be accurate within a few percent.
This shows that describing electron scattering even from a singly ionized target is  
significantly easier than the corresponding problem of collisions with a neutral hydrogen
atom \citep{BHSBB1996}.

Figure~\ref{fig:fig4} shows the QDT results for elastic electron collisions with He$^+$ and 
selected other one-electron
ions up to Fe$^{25+}$. In this case, the quantum defects were taken from the database
of the Opacity Project \citep{OP}.  Even though it is difficult to accurately read numbers
off their graphs, there are significant deviations between our results and
those of \citet{AMG1989}.  While this may be understandable for the He$^+$ target, one
would expect much better agreement for the highly charged ions.  


\begin{figure}[t]
\centerline{\includegraphics[width=0.80\textwidth]{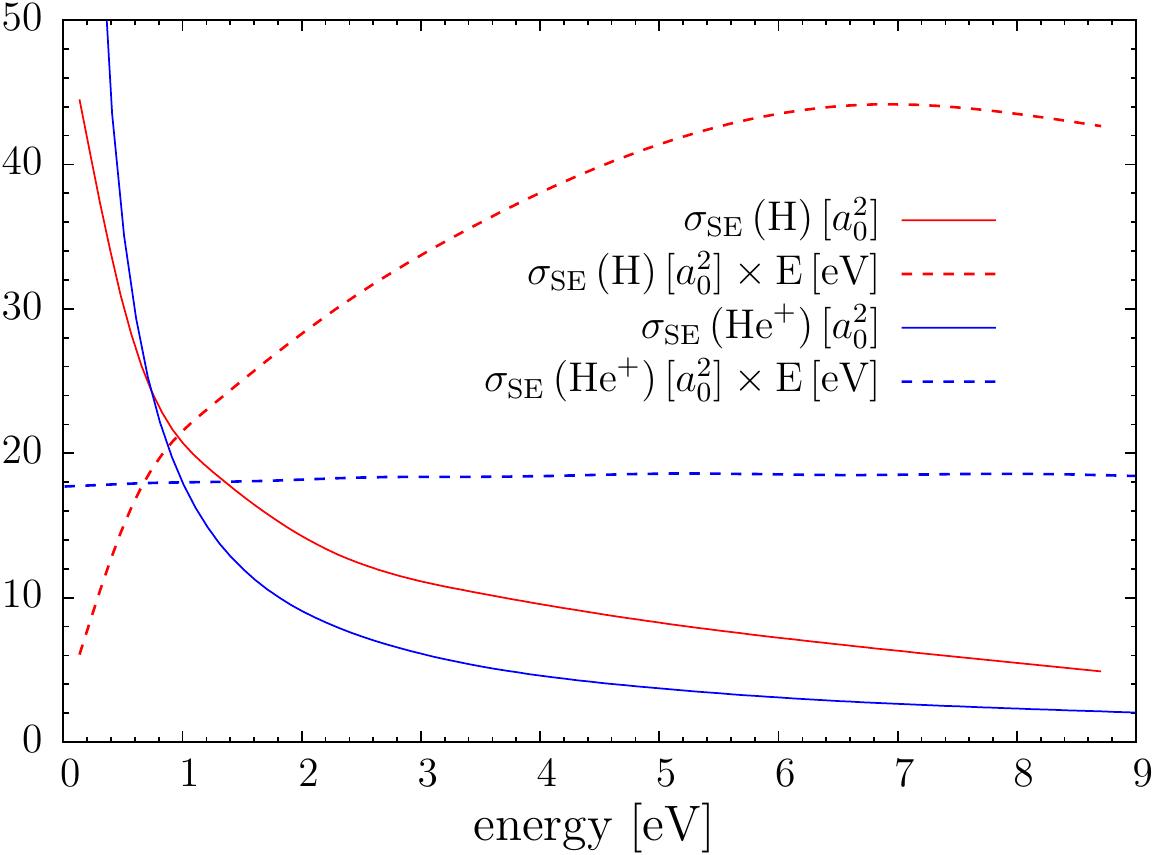}}
\caption{\label{fig:compare} Spin-exchange cross sections for He$^+$ (obtained from the RM-23 model)
and neutral atomic hydrogen.  The latter were obtained using phase shifts from \citet{Schwartz1961} 
for the \hbox{s-wave} and \citet{Bhatia2004} for the \hbox{p-wave}.  Also shown are the cross sections multiplied by the
energy (in eV), in order to show the entirely different energy dependence, in particular at very low energies.
See the electronic edition of the Journal for a color version of this figure.}
\end{figure}

We finish with a comparison of the SE cross section for neutral hydrogen and
He$^+$.  The predictions are shown in Fig.~\ref{fig:compare} for energies below the
first negative-ion resonance in e$-$H collisions. The hydrogen results were obtained 
using the highly accurate phase shifts of \citet{Schwartz1961} for the \hbox{s-wave} and
\citet{Bhatia2004} for the \hbox{p-wave}. 
[Similar results, using slightly different phaseshifts, were published by \citet{FF2007}.  Note, however, that their definition of the spin-exchange 
cross section (see Eq.~10 of the paper) differs from ours in Eq.~(\ref{eq:sigmaex}) above by a factor of~4.]

In addition to plotting the SE cross section, we
also show it multiplied by the energy (taken in eV in order to ensure good visibility on the graph).
Due to the QDT scaling for an ionic target such as He$^+$, the cross section grows proportional to $1/E$
for $E\to 0$.  For neutral hydrogen, on the other hand, low-energy collisions are dominated by
the polarization potential.  The threshold behavior is very different, as seen most clearly
when the cross section is multiplied by the energy. Specifically, the spin-exchange cross section
for neutral hydrogen will approach the finite value of
\begin{equation}
  \lim_{E \to 0} \sigma_{\rm SE}(E) = \pi (a^t - a^s)^2,
\end{equation}  
where $a^t$ and $a^s$ are the scattering lengths for triplet and singlet scattering, respectively.

Finally, the collisional rate coefficients are:
\begin{equation}
  K(T) = \sqrt{\frac{8 k_B T}{\pi m}} \int_0^\infty \sigma(\bar{E}) \; {\rm e}^{-\bar{E}}\, d\bar{E}, 
\end{equation}  
where $m$ is the electron mass, $T$ is the temperature in K, $k_B$ is the Boltzmann constant, 
and $\bar{E} \equiv E/(k_B T)$.  Given the $1/E$ dependence of the SE cross section 
for e$-$He$^+$ collisions, with a slope of 17.3~$a_0^2 \times {\rm eV}$, we obtain the explicit form
\begin{equation}
  K(T) = 3.49 \times 10^{-6} \,T^{-1/2} \, \rm cm^3/s. 
\end{equation}
This yields, for example, $K(T=10\,{\rm K}) = 1.1 \times 10^{-6}\,\rm cm^3/s$ 
and $K(T=100\,{\rm K}) = 3.49 \times 10^{-7}\,\rm cm^3/s$. The enormous rate coefficients in the 
scattering of electrons from \HeII\ are due to the fact that the SE cross section scales as $1/E$ and
hence is infinite at zero energy.

\section{Summary}
We have presented short-range phase shifts and spin-exchange cross sections for
elastic electron scattering from selected one-electron ions from He$^+$ to Fe$^{25+}$.
The results were explained in a simple physical picture based on quantum defect theory.
In the low-energy range, which is of particular interest for astrophysical applications,
it seems unnecessary to perform explicit numerical calculations for the scattering
process, as long as accurate quantum defects are available. Our results differ 
significantly from those published previously, partly due to the more sophisticated
approach used in the present work (for small~$Z$) and partly due to a possible 
plotting error in~\citet{AMG1989}. The entirely different energy dependence of the 
spin-exchange cross section for ionic versus neutral targets leads to very large
rate coefficients for the electron-induced 3.46~cm line transition in \HeII\ at low temperatures.

\section*{Acknowledgments}
This work was supported by the United States National Science Foundation
under grant No.~PHY-1068140 (KB) and through a grant to the Institute for Theoretical
Atomic, Molecular and Optical Physics at Harvard University and the Harvard-Smithsonian
Center for Astrophysics.



\end{document}